# Inhomogeneous superconductivity induced by interstitial Fe deintercalation in oxidizing-agent-annealed and HNO$_3$-treated Fe$_{1+y}$(Te$_{1-x}$Se$_x$)

J. Hu, G.C. Wang, B. Qian, and Z.Q. Mao*

Department of Physics and Engineering Physics, Tulane University, New Orleans,

Louisiana 70118, USA

Abstract

We have systematically investigated the annealing effect on the superconductivity of iron chalcogenide Fe$_{1+y}$(Te$_{1-x}$Se$_x$). The atmospheres used for annealing include O$_2$, N$_2$, I$_2$ vapor, air and vacuum. We observed that annealing in O$_2$, I$_2$ and air could enhance superconductivity for the underdoped samples, consistent with the results reported in literatures. Interestingly, we found that annealing in N$_2$ also leads to superconductivity enhancement, similar to the annealing effects of O$_2$, I$_2$ and air. However, vacuum annealing does not enhance superconductivity, which indicates that the enhanced superconductivity in O$_2$-, N$_2$- , I$_2$- and air-annealed samples is not due to improved homogeneity. In addition, we have treated the underdoped samples with nitric acid, which is found to enhance superconductivity as well. Our analyses of these results support the argument that the superconductivity enhancement, caused either by annealing or nitric acid treatment, originates from the variation of interstitial Fe. The interstitial Fe, which is destructive to superconducting pairing, can be reduced by annealing in oxidation agents or nitric acid treatment. We also find that although N$_2$-, O$_2$- and air-annealed samples exhibit strong superconducting diamagnetism with -4$\pi\chi$ ~1 ($\chi$, dc magnetic susceptibility) for some samples, their actual superconducting volume fraction probed by specific heat is low, ranging from 10% to 30% for 0.09 < $x$ < 0.3, indicating that the superconductivity suppression remains significant even in annealed samples. The strong diamagnetism is associated with the superconducting shielding effect on the non-superconducting phase. We have also established the phase diagram of the annealed samples and compared it with that of the as-grown samples. The effect of annealing on the interplay between magnetism and superconductivity is discussed.

zmao@tulane.edu

The iron chalcogenide superconductor $Fe_{1+y}(Te_{1-x}Se_x)$ has attracted a great deal of interest due to its unique properties. While this system possesses the simplest structure among Fe-based superconductors, it exhibits a complex phase diagram compared to iron pnictide superconductors: an unusual intermediate phase characterized by charge carrier weak localization ($0.09 < x < 0.29$, noted as Region II in the phase diagram reported in ref. [1]) lies between the antiferromagnetic metallic phase (x < 0.09, Region I) and the bulk superconducting phase ($x > 0.29$, Region III) [1-2]. This contrasts with iron pnictide superconductors in which superconductivity occurs immediately following the suppression of long range AFM order [3-4] or coexists with the antiferromagnetism within a certain composition region [5-8]. Such difference between these two systems is attributed to the fact that in iron chalcogenides there exists a coexistence of two competing magnetic correlations at $(\pi,0)$ and $(\pi,\pi)$, while iron pnictides are characterized by only $(\pi,\pi)$ magnetic correlations. Se substitution for Te in $Fe_{1+y}(Te_{1-x}Se_x)$ tunes the relative strength of two magnetic correlations [1]; the $(\pi,0)$ magnetic correlations are significantly weakened from the underdoped to optimally doped region, whereas the $(\pi,\pi)$ magnetic correlations strengthen accordingly. Our previous work has revealed that the weakly localized state in the underdoped region is associated with the $(\pi,0)$ magnetic fluctuations. Strong evidence for the incoherent magnetic scattering of charge carriers by the $(\pi,0)$ magnetic fluctuations has been observed [2]; such incoherent magnetic scattering leads to considerable electronic disorders, thus resulting in weak charge carrier localization and superconductivity suppression in the underdoped region [2].

Another remarkable characteristic of $Fe_{1+y}(Te_{1-x}Se_x)$ is that both the magnetism and superconductivity of this system are sensitively dependent on the Fe non-stoichiometry, which originates from the partial occupation of excess Fe at the interstitial sites of the Te/Se layer [9]. For the undoped parent compound $Fe_{1+y}Te$, its AFM wave vector can be tuned by the excess Fe, changing from commensurate to incommensurate when $y$ is increased above 0.076 [9]. For optimally doped samples, the increase of interstitial Fe was found to suppress superconductivity and cause charge carrier localization [10]. Such superconductivity suppression is attributed to the enhanced $(\pi,0)$ magnetic fluctuations by interstitial Fe [1].

In addition, $Fe_{1+y}(Te_{1-x}Se_x)$ exhibits an intriguing annealing effect. The underdoped samples with $0.09 \leq x < 0.3$, which show only a trace of superconductivity in as-grown single crystals [1-2, 11], were reported to display noticeably enhanced superconductivity after being annealed in air [12], oxygen [13], vacuum [14-15] and $I_2$ vapor [16]. The interpretations for such annealing effects are inconsistent in literatures, including the improved homogeneity [14-15], the oxygen intercalation [13] and the deintercalation of interstitial Fe [12, 16]. In order to address this issue, we have performed comprehensive annealing-effect studies on $Fe_{1+y}(Te_{1-x}Se_x)$ single crystals using various annealing atmospheres, including vacuum, oxygen ($O_2$), nitrogen ($N_2$), iodine ($I_2$) vapor and air. We have also investigated the superconductivity of this system using the samples treated with nitric acid $HNO_3$. We found that the samples annealed in $O_2$, $N_2$, iodine vapor and air, as well as the samples treated with $HNO_3$, all show clear superconductivity enhancement, consistent with the results reported in literatures [12-16] (Note that $N_2$ annealing effect was not previously studied). In contrast, we did not observe any superconductivity enhancement in vacuum-annealed samples. Our detailed analyses of these results provide strong support for the argument that the enhanced superconductivity in the annealed samples coms from the deintercalaion of interstitial Fe. The superconductivity enhancement in $HNO_3$-treated samples can be attributed to a similar mechanism. Moreover, from our specific heat measurements, we also found that the actual superconducting volume fraction $V_{SC}$ of annealed, underdoped samples ($0.09 \leq x < 0.3$) remains as low as 10-30% though their magnetic susceptibility data display strong superconducting diamagnetism with $0.6 < -4\pi\chi < 1$. This indicates that in underdoped samples annealing induces considerable inhomogeneous superconductivity and that the superconductivity suppression by the $(\pi,0)$ magnetic fluctuations still remains dominant. We have established the phase diagram of the $N_2$-annealed samples and discussed the effect of annealing on the interplay between magnetism and superconductivity.

$Fe_{1+y}(Te_{1-x}Se_x)$ single crystals used in this study were synthesized using a flux method [10] and were shown to be tetragonal phase with the space group *P*4/*nmm* at room temperature by X-ray diffraction measurements. Since the superconductivity of $Fe_{1+y}(Te_{1-x}Se_x)$ is sensitive to Fe non-stoichiometry as noted above [10, 17], two types of samples were used for annealing in various atmospheres (including $O_2$, $N_2$, $I_2$ vapor, air and vacuum) and $HNO_3$ treatment, with one type having less excess Fe ($y \sim 0.02$) and the other having rich excess Fe ($y \sim 0.14$). The amount of excess Fe was examined by an energy dispersive X-ray spectrometer (EDXS). The dc magnetic susceptibility was measured using a superconducting quantum interference device (SQUID, Quantum Design) under magnetic field of 30 Oe with zero-field cooling history. The specific heat was measured with an adiabatic relaxation technique using a physical property measurement system (PPMS, Quantum Design).

For $O_2$ annealing, we have selected a group of single crystals with typical compositions with $x$ = 0.05, 0.1, 0.15, 0.20, 0.25, 0.30, 0.35, 0.45 and $y$~0.02. These samples were sealed in quartz tubes filled with 0.5 atm ultra-high purity $O_2$ gas after the tubes were pumped to high vacuum (~$10^{-5}$ torr), and then were annealed at 300 ºC for 10 hours. This annealing condition was found to be optimal from our trials and errors; the samples annealed with this condition have the strongest diamagnetism for a given composition. We also found that longer-time annealing leads the samples to degrade. Figure 1a presents the magnetic susceptibility $\chi$ data of these samples. Except for the $x$ = 0.05 sample, all the annealed, underdoped samples with $x$ = 0.1 - 0.35 exhibit remarkably enhanced superconductivity with respect to the as-grown samples within the same composition range. This can be seen clearly from their increased superconducting diamagnetism, as manifested in their large values of -4$\pi\chi$ ( = 0.5 - 1). This result is consistent with the earlier report on $O_2$ annealing effects [13]. Such strong diamagnetism of annealed samples are in sharp contrast with the weak diamagnetism observed in the as-grown samples where -4$\pi\chi$ < 0.03 for $x$= 0.1-0.3 [1, 11]. The significantly increased diamagnetism seemingly implies the presence of bulk superconductivity in underdoped samples. However, as we will show below, the actual superconducting volume fraction probed in the specific heat measurements is far less than the $V_{SC}$ value given by -4$\pi\chi$. For example, the $O_2$-annealed sample with $x = 0.2$ has the actual superconducting volume faction of 16% though it exhibits ideal diamagnetism with $-4\pi\chi \sim 1$. The strong diamagnetism results from the superconducting shielding effect on the non-superconducting phase, as discussed below. For the optimally doped sample with $x = 0.45$, $O_2$-annealing hardly has any effect on superconductivity compared to the underdoped samples.

For comparison with the $O_2$ annealing effect, we have annealed another group of samples with $x = 0.05 - 0.45$ in ultra-high purity $N_2$ using the same conditions as used for $O_2$ annealing, *i.e.* 300 ºC for 10 hours. Surprisingly, we found that the $N_2$-annealed, underdoped samples with $x = 0.1 - 0.35$ display superconductivity enhancement similar to that seen in $O_2$-annealed samples, as shown in Figure 1b where the magnetic susceptibility data for $N_2$-anealed samples are presented.  In contrast, $N_2$-annealing does not generate any noticeable superconductivity enhancement for the $x = 0.05$ sample with the AFM order and hardly has any effect on the bulk superconductivity of the $x = 0.45$ optimally-doped sample.

 In general, annealing can increase chemical homogeneity for alloy systems. A natural question here is whether the superconductivity enhancement observed in $O_2$ and $N_2$ annealed samples is associated with the improved homogeneity; this has actually been proposed as one possible origin [14-15], as noted above. To address this question, we have conducted "in-situ" vacuum annealing at 400 ºC for underdoped samples. The samples were annealed immediately after the crystal growth without being taken out from the sealed quartz tube (*i.e.* the entire batch of crystals was annealed). Since the samples were not exposed to any atmospheres in this process, the contamination is minimized. The annealing process lasted for 20 days, which should significantly improve homogeneity. However, as shown in the inset of Figure 1a, such long-time "in-situ" annealing does not produces any sizable effect on superconductivity with respect to the as-grown samples; the superconducting diamagnetism remains very weak with $-4\pi\chi \ll 0.01$ for all annealed samples. This observation clearly indicates that the remarkably enhanced superconductivity in any annealed $Fe_{1+y}(Te_{1-x}Se_x)$ samples cannot be attributed to increased homogeneity. We note that our results are inconsistent with the previous studies by Noji *et al*. who reported that superconductivity could be greatly enhanced by vacuum annealing at 400 ºC for 100 or 200 hours [14-15]. Such inconsistency is probably due to the contamination by $O_2$ or $N_2$ during the annealing process. We also tested some samples which were annealed in various vacuum pressures and found that annealing could enhance superconductivity when the vacuum pressure is $\geq 10^{-3}$ torr (data not shown here). This implies that high vacuum annealing is necessary to examine the effect of improved homogeneity on superconductivity.

Given that the improved homogeneity by annealing is not responsible for the superconductivity enhancement, the observed enhancement of superconductivity in $O_2$- and $N_2$- annealed samples should be attributed to either the variation of interstitial Fe or chemical intercalation. Both possibilities have been proposed in literatures [12-13], as indicated above. For $O_2$-annealed samples, Kawasaki *et al*. speculated that $O_2$ may be intercalated between layers, which would generate additional holes to compensate the electrons contributed by interstitial Fe [13]. If this were the case, we would expect to observe a difference in superconductivity between $O_2$- and $N_2$- annealed samples since oxygen and nitrogen have different chemical valences; and annealing should also have an effect on optimally doped samples in this case. Our observation of the similar superconductivity enhancement of $O_2$- and $N_2$- annealed samples, as well as the absence of annealing effect on optimally doped samples, points to the scenario that the annealing induced superconductivity enhancement originates from the variation of interstitial Fe, as suggested in ref. [12]. Since both $O_2$ and $N_2$ are chemically active for Fe, annealing $Fe_{1+y}(Te_{1-x}Se_x)$ samples in $O_2$ and $N_2$ is most likely to remove the interstitial Fe to some extent via oxidation reaction. Since the interstitial Fe is in favor of stabilizing the $(\pi,0)$ magnetic fluctuations which are destructive to superconducting pairing as mentioned above [1-2], the reduction of interstitial Fe would naturally enhance superconductivity.

To further demonstrate the idea that an oxidation agent is capable of extracting interstitial Fe, we have treated the $x = 0.20$ as-grown samples with iodine vapor and nitric acid. Both $I_2$ and $HNO_3$ are well known as strong oxidation agents. The $I_2$-vapor annealing has actually been reported to be effective to reduce interstitial Fe and enhance superconductivity [16, 18]. Our observation of superconductivity enhancement in the underdoped samples annealed in $I_2$-vapor, as shown in Figure 2a, is consistent with the previously reported results [16]. We also found that long-time annealing ( > 10 hrs) in $I_2$ could significantly degrade the single crystal samples. The $HNO_3$ treatment was preceded by immersing the sample in the $HNO_3$ solution with the concentration of 2 N. An identical sample was used for sequential immersions with various periods of time; the magnetic susceptibility of the sample was measured following each period of immersion. As shown in Figure 2b, as the accumulated immersion time increases, the superconductivity remarkably enhances; the superconducting diamagnetism increases by almost 20% as the immersion time extends to 1.5 hrs. Further increase of immersion time, however, leads the sample to degrade. These results strongly support that an oxidation agent can indeed lead to deintercalation of interstitial Fe, which, from another aspect, suggests that the superconductivity enhancement induced by annealing in $O_2$, $N_2$ or air, should all be driven by a similar mechanism, i.e. the deintercalation of interstitial Fe. The same mechanism should also be applicable to the superconductivity enhancement induced by annealing in the $Fe_{1+y}(Te_{1-x}S_x)$ system [19-20]. We notice that Deguchi *et al*. recently reported that alcoholic beverages, such as red wine, white wine, and beer can enhance the superconductivity of $Fe_{1+y}Te_{1-x}S_x$ [21]. Such superconductivity enhancement is ascribed to the deintercalation of interstitial Fe by the organic acids included in the alcoholic beverages [22]. Our observation of superconductivity enhancement in $HNO_3$-treated samples is in good agreement with their claim.

To seek further evidence for the interstitial Fe deintercalation induced by annealing, we have also synthesized the sample with nominal composition $Fe_{1.14}Te_{0.6}Se_{0.4}$. While the Se content in this sample is close to optimal doping, it does not show bulk superconductivity owing to rich interstitial Fe (see Figure 3), consistent with our previous reports [10]. We have annealed the crystals with this composition in various atmospheres, including vacuum, $N_2$, $O_2$ and $I_2$ vapor. Like the vacuum-annealed, underdoped samples discussed above, the vacuum-annealed $Fe_{1.14}Te_{0.6}Se_{0.4}$ sample also does not show any observable superconductivity enhancement. However, annealing in O$_2$, N$_2$, and iodine vapor, as well as HNO$_3$ treatment, for this sample can all enhance superconductivity more or less, as shown in Figure 3. Since the superconductivity of the optimally doped samples with less interstitial Fe ($y \sim 0.02$) does not exhibit any annealing effect, as described above, the enhanced superconductivity in Fe$_{1.14}$Te$_{0.6}$Se$_{0.4}$ by either annealing or HNO$_3$ treatment is precisely consistent with the argument that interstitial Fe is the key tuning parameter for the various annealing effects as well as HNO$_3$ treatment.

As stated above, the iron chalcogenide Fe$_{1+y}$(Te$_{1-x}$Se$_x$) system exhibits a unique phase diagram. Our phase diagram reported in ref. [1-2] is based on as-grown crystals. One of the salient features of that phase diagram is the presence of a weakly localized phase in the underdoped region ($0.09 \leq x < 0.3$) where only a trace of superconductivity is observed. Since annealing in an oxidation agent can enhance the superconductivity of underdoped samples as discussed above, it would be interesting to examine how the phase diagram of annealed samples differs from that of as-grown samples. We notice that several groups have established the phase diagram using samples annealed in air [12], vacuum [14] and O$_2$ [13]. Those phase diagrams show that bulk superconductivity occurs immediately following the suppression of the long range AFM order [13-14] or could coexists with the AFM order [12]. However, in those phase diagrams, the determination of bulk superconductivity was based on the susceptibility measurements. This approach could become inaccurate for several reasons. In particular, when the sample involves non-superconducting phase, which is the case for the annealed, underdoped Fe$_{1+y}$(Te$_{1-x}$Se$_x$) samples as we will show below, the superconducting volume fraction estimated by - $4\pi\chi$ becomes unreliable. This is because that as the sample includes non-superconducting phases or voids, the superconducting phase may shield non-superconducting phases, resulting in overestimate of superconducting volume fraction; this has been well addressed in literature [23]. In some cases, the overestimate of $V_{SC}$ caused by the shielding effect is huge, as indicated above. Additionally, the sample-shape dependent demagnetization effect may also cause ambiguity in the $V_{SC}$ estimated by -$4\pi\chi$ and it is usually difficult to obtain the demagnetization factor to make a correction.

The specific heat measurement is well known to be the most precise approach for evaluating superconducting condensation; electronic specific heat should vanish at the zero temperature limit when perfect bulk superconducting state is achieved (*i.e.* $V_{SC}$ = 100%). Therefore $V_{SC}$ can be estimated by measuring the residual electronic specific heat. We performed systematic specific heat measurements on the $N_2$-annealed samples. The specific heat data of the $N_2$-annealed samples are presented in Figure 4. Although the as-grown samples with $0.09 \leq x < 0.3$ do not exhibit any features associated with superconductivity in specific heat [1], we observed superconducting anomalies near $T_c$ in the specific heat data of $N_2$-annelaed samples in this composition range, confirming the superconductivity enhancement suggested by the strong diamagnetism (see Figure 1b). However, the superconducting transitions of these samples are fairly broad compared to the optimally doped sample as shown in Figure 4b, indicating considerably inhomogeneous superconductivity. Moreover, the $V_{SC}$ estimated from residual electronic specific heat for these samples remains small as shown in the following analyses. For the $x$ = 0.05, annealed sample, we observed an anomaly associated the AFM transition in its specific heat data (see the inset in Figure 4a), but did not see any sign of superconductivity, consistent with its susceptibility data shown in Figure 1b. This result excludes the possibility that annealing could lead to the coexistence of bulk superconductivity and long-range AFM order. For the optimally-doped, annealed sample ($x = 0.45$), the superconducting anomaly peak of the specific heat is nearly identical to that seen in the as-grown sample and the only annealing effect is reflected in the slight decrease of residual electronic specific heat (see below).

As shown in the insets of Figure 4a and 4b, the residual electronic specific heat coefficient $\gamma_{res}$ can be derived by fitting the specific heat data using $C = \gamma_{res}T + \beta T^3$ at temperatures well below $T_c$, where $\gamma_{res}T$ and $\beta T^3$ represent the residual electronic specific heat and the phonon specific heat respectively. $\gamma_{res}$ obtained from fitting ranges from 56 mJ/mol K$^2$ to 23 mJ/mol K$^2$ for $0.1 \leq x \leq 0.3$ and decreases with an increase of Se content (see Figure 5b). These values of $\gamma_{res}$ are much greater than that of the optimally doped sample with $x = 0.45$ ($\gamma_{res}$ =1.53 mJ/mol K$^2$), indicating that the annealing-induced superconductivity in the underdoped region is far from the bulk phenomenon. This observation is consistent with the previous results of specific heat measurements on O$_2$-annealed samples, which also show large $\gamma_{res}$ in the underdoped region [15].

In general, the exact superconducting volume faction $V_{SC}$ can be derived from $\gamma_{res}$, i.e. $V_{SC} = 1 - \gamma_{res} / \gamma$, where $\gamma$ is the normal state Sommerfeld coefficient. However, since the iron chalcogenide superconductors possess very high upper critical fields [24-27], as do the iron pnictide superconductors [28-32], it is difficult to separate the electronic specific heat from the phonon contribution through measurements on the normal state achieved by applying a magnetic field. To obtain the electronic specific heat, we have employed the non-superconducting samples as reference to separate electronic specific heat from phonon contribution; this approach has been shown to be very effective as addressed in our early work [2, 33] as well as in iron pnictides [34]. We used the as-grown single crystals as references for the specific heat data analyses of the $x$ = 0.1 - 0.25 samples and Cu-doped, non-superconducting samples as references for the $x$=0.3, 0.35 and 0.45 samples (see ref.[33] for details). From these analyses, we obtained $\gamma$ for all $N_2$-annealed samples, as shown in Figure 5b where we also present $\gamma$ for as-grown samples, which was obtained in our previous work [2], for comparison.

From $\gamma$ and $\gamma_{res}$ shown in Figure 5b, we estimated $V_{SC}$ (= $1 - \gamma_{res} / \gamma$) for all annealed samples, which is presented using a contour plot in the phase diagram in Figure 5a where the AFM phase boundary denoted by $T_N$ was determined in our previous work [1-2] and $T_c^{annealed}$ stands for the superconducting anomaly peak temperature in the specific heat of annealed samples. The evolution of $V_{SC}$ in the underdoped region shows a remarkable difference between as-grown and annealed samples. The as-grown samples show only a trace of superconductivity, with $V_{SC}$ < 3%, for 0.09 < $x$ < 0.3 [1-2]. However, for the annealed samples in this composition region, $V_{SC}$ rises to 10%-30%. $V_{SC}$ steeply increase up to > 90% near $x$ = 0.4. The low superconducting volume fraction in the underdoped region clearly indicates that the superconductivity suppression in this region remains dominant even in annealed samples. Due to the existence of a non-superconducting phase in these samples, the superconducting shielding effect on the non-superconducting phase as mentioned above is naturally expected, which explains the overestimate of $V_{SC}$ by $-4\pi\chi$. For example, the overestimate exceed 50 - 60% for the $x =$ 0.1 and 0.2 samples (see Figure 1b and Figure 5a). A similar shielding effect also occurs in the underdoped samples annealed in $O_2$. We measured the specific heat of the $O_2$-annealed sample with $x = 0.2$ (see Figure 4b). Although this sample exhibits the strongest diamagnetism, with $-4\pi\chi \sim 1$, its $\gamma_{res}$ is found to be ~47 mJ/mol K$^2$ (see the inset of Figure 4b), with $V_{SC} \approx 16\%$. In addition, we have examined the air annealing effect by specific heat measurements on an $x = 0.1$ sample annealed in air 270 °C for 2 hrs. This annealing condition is the same as that used by Dong *et al.* [12] who established the phase diagram of air-annealed samples using resistivity and susceptibility data and claimed that bulk superconductivity coexist with the AFM order in air-annealed samples for $0.05 \leq x \leq 0.18$. Our air-annealed sample exhibits superconducting diamagnetism comparable to that reported in ref. [12], with $-4\pi\chi \sim 0.35$ (see the inset of Figure 1b). However, the specific heat data of this sample does not reveal bulk superconductivity (see the left inset of Figure 4b), with $\gamma_{res} = 58$ mJ/mol K$^2$ and $V_{SC}$ less than 10%.

Given that the superconductivity of our current $N_2$- and $O_2$- annealed samples is far from a perfect bulk phenomenon for $0.09 < x < 0.3$, a natural question is whether it is possible to increase $V_{SC}$ up to 100% by optimizing annealing conditions. The answer to this question is most likely "No" for the following reasons. First, we have made our best efforts to optimize the annealing conditions. We have annealed samples in both $O_2$ and $N_2$ with various conditions; both annealing temperatures and time have been tuned to obtain the samples having the strongest superconducting diamagnetism. The $O_2$- and $N_2$-annealed samples used in this study were all annealed with the optimized conditions, as noted above. Secondly, Fe non-stoichiometry was found to be dependent on Se content in iron chalcogenides [11, 35]. The minimal content of interstitial Fe $y_{min}$ required to stabilize the crystal structure likely decreases with an increase of Se content and becomes small for $x \geq 0.4$ where $V_{SC} > 90\%$ in as-grown samples. Thus, for underdoped samples, it is difficult to reduce interstitial Fe to a level as low as that in optimally doped samples through annealing in an oxidation agent. This also explains why the steep increase of $V_{SC}$ up to $> 90\%$ takes place near $x = 0.4$ in both phase diagrams of as-grown and annealed samples. For the phase diagram of as-grown samples in ref. [1-2], although we intentionally selected samples with less excess Fe ($y\sim0.02$) for the entire composition region, the determination of excess Fe was based on EDXS measurements, which have limited resolution. The actual content of excess Fe could differ from that measured by EDXS and the actual difference of excess Fe content between underdoped and optimally doped samples may be within the error bar of EDXS.

Finally, let's examine the difference in the evolution of Sommerfeld coefficient $\gamma$ between as-grown and annealed samples (see Figure 5b). For as-grown samples, $\gamma$ appears to be a key tuning parameter for superconductivity; it shows large values (57-65 mJ/mol K$^2$) in the underdoped region where superconductivity is suppressed and drops steeply for $x > 0.35$ where bulk superconductivity develops. For $N_2$-annealed samples, the variation of $\gamma$ also couples with the evolution of superconductivity; $V_{SC}$ gradually increases from 10% to 44% as $\gamma$ decreases almost linearly from 62.3 mJ/mol K$^2$ for $x$ = 0.1 to 32 mJ/mol K$^2$ $x$ = 0.35. $\gamma$ drops to 26.5 mJ/mol K$^2$ for optimally doped samples whose $V_{SC}$ > 90%. From this comparison, it can be seen that annealing reduces $\gamma$, *i.e.* electronic specific heat for underdoped samples. As indicated above, the large $\gamma$ values for as-grown, underdoped samples are attributed to enhanced electronic disorders arising from incoherent magnetic scattering by ($\pi$,0) magnetic fluctuations [2]; such electronic disorders result in weak charge carrier localization and superconducting pair breaking. For annealed samples, interstitial Fe is partially deintercalated in the annealing process as discussed above. Since interstitial Fe favors stabilizing ($\pi$,0) magnetic correlation [1], the reduction of interstitial Fe via annealing naturally weakens ($\pi$,0) magnetic fluctuations. Therefore annealed samples should have weaker ($\pi$,0) magnetic fluctuations than as-gown samples in the underdoped region. Under this circumstance, the superconductivity enhancement by annealing can easily be understood.

In summary, we have systematically studied the superconductivity of iron chalcogenide Fe$_{1+y}$(Te$_{1-x}$Se$_x$) single crystals annealed in various atmospheres (including vacuum, O$_2$, N$_2$, I$_2$ vapor and air) and treated with HNO$_3$. We found that the superconductivity of the underdoped samples with 0.09< $x$ < 0.3, which behaves as non-bulk behavior ($V_{SC}$<3%) in the as-grown single crystals, can be enhanced not only by O$_2$-, I$_2$- and air-annealing as reported in literatures, but also by N$_2$-annealing or HNO$_3$ treatment. However, annealing in a vacuum does not have any effect on superconductivity. From analyses of these results, we have demonstrated that the superconductivity enhancement induced by annealing or HNO$_3$ treatment originates from interstitial Fe deintercalation, rather than from improved homogeneity or oxygen/nitrogen intercalation. Our systematic specific heat measurements revealed that the superconductivity enhanced by annealing is considerably inhomogeneous and that the superconducting volume fraction is as low as 10%-30% for $0.09 < x < 0.3$. However, the superconducting diamagnetism of these samples is strong, with $0.5 < -4\pi\chi \leq 1$, due to the superconducting shielding effect on the non-superconducting phase. The large residual electronic specific heat of the annealed, underdoped samples, together with their large normal state Sommerfeld coefficients, suggest that the pair-breaking by $(\pi,0)$ magnetic fluctuations remain significant in the underdoped region. Therefore, the phase diagram of annealed samples bears significant similarity to that of as-grown samples.

ACKNOWLEDGMENT

This work is supported by the NSF under grant DMR-1205469 and the LA-SiGMA program under Award No. EPS-1003897 -1-0031

[1]     Liu T J *et al.* 2010 *Nat. Mater.* **9** 718

[2]     Hu J, Liu T J, Qian B, and Mao Z Q 2011 *arXiv:1111.0699*

[3]     Zhao J *et al.* 2008 *Nat. Mater.* **7** 953

[4]     Luetkens H *et al.* 2009 *Nat. Mater.* **8** 305

[5]     Drew A J *et al.* 2009 *Nat. Mater.* **8** 310

[6]     Chen H *et al.* 2009 *Europhys. Lett.* **85** 17006

[7]     Chu J-H, Analytis J G, Kucharczyk C, and Fisher I R 2009 *Phys. Rev. B* **79** 014506

[8]     Nandi S *et al.* 2010 *Phys. Rev. Lett* **104** 057006

[9]     Bao W *et al.* 2009 *Phys. Rev. Lett* **102** 247001

[10]    Liu T J *et al.* 2009 *Phys. Rev. B* **80** 174509

[11]    Sales B C, Sefat A S, McGuire M A, Jin R Y, Mandrus D, and Mozharivskyj Y 2009 *Phys. Rev. B* **79** 094521

[12]    Dong C, Wang H, Li Z, Chen J, Yuan H Q, and Fang M 2011 *Phys. Rev. B* **84** 224506

[13]    Kawasaki Y, Deguchi K, Demura S, Watanabe T, Okazaki H, Ozaki T, Yamaguchi T, Takeya H, and Takano Y 2012 *Solid State Commun.* **152**, 1135

[14]    Noji T, Suzuki T, Abe H, Adachi T, Kato M, and Koike Y 2010 *J. Phys. Soc. Jpn.* **79** 084711

[15]    Noji T, Imaizumi M, Suzuki T, Adachi T, Kato M, and Koike Y 2011 J. Phys. Soc. Jpn. **81**, 054708

[16]    Rodriguez E E, Stock C, Hsieh P-Y, Butch N P, Paglione J, and Green M A 2011 *Chem. Sci.* **2** 1782

[17]    McQueen T M *et al.* 2009 *Phys. Rev. B* **79** 014522

[18]    Rodriguez E E, Zavalij P, Hsieh P-Y, and Green M A 2010 *J. Am. Chem. Soc.* **132** 10006

[19]    Mizuguchi Y, Deguchi K, Tsuda S, Yamaguchi T, and Takano Y 2010 *Europhys. Lett.* **90** 57002

[20]     Mizuguchi Y, Deguchi K, Kawasaki Y, Ozaki T, Nagao M, Tsuda S, Yamaguchi T, and Takano Y 2011 *J. Appl. Phys.* **109** 013914

[21]     Deguchi K *et al.* 2012 *arXiv:1203.4503*

[22]     Deguchi K *et al.* 2012 *Supercond. Sci. Technol.* **24** 055008

[23]     Ando Y, and Akita S 1990 *Jpn. J. Appl. Phys.* **29** L770

[24]     Kida T, Matsunaga T, Hagiwara M, Mizuguchi Y, Takano Y, and Kindo K 2009 *J. Phys. Soc. Jpn.* **78** 113701

[25]     Fang M, Yang J, Balakirev F F, Kohama Y, Singleton J, Qian B, Mao Z Q, Wang H, and Yuan H Q 2010 *Phys. Rev. B* **81** 020509

[26]     Khim S, Kim J W, Choi E S, Bang Y, Nohara M, Takagi H, and Kim K H 2010 *Phys. Rev. B* **81** 184511

[27]     Lei H, Hu R, Choi E S, Warren J B, and Petrovic C 2010 *Phys. Rev. B* **81** 094518

[28]     Hunte F *et al.* 2008 *Nature* **453** 903

[29]     Jaroszynski J *et al.* 2008 *Phys. Rev. B* **78** 174523

[30]     Altarawneh M M, Collar K, Mielke C H, Ni N, Bud'ko S L, and Canfield P C 2008 *Phys. Rev. B* **78** 220505

[31]     Yuan H Q, Singleton J, Balakirev F F, Baily S A, Chen G F, Luo J L, and Wang N L 2009 *Nature* **457** 565

[32]     Baily S A, Kohama Y, Hiramatsu H, Maiorov B, Balakirev F F, Hirano M, and Hosono H 2009 *Phys. Rev. Lett* **102** 117004

[33]     Hu J, Liu T J, Qian B, Rotaru A, Spinu L, and Mao Z Q 2011 *Phys. Rev. B* **83** 134521

[34]     Popovich P, Boris A V, Dolgov O V, Golubov A A, Sun D L, Lin C T, Kremer R K, and Keimer B 2010 *Phys. Rev. Lett* **105** 027003

[35]     Zajdel P, Hsieh P-Y, Rodriguez E E, Butch N P, Magill J D, Paglione J, Zavalij P, Suchomel M R, and Green M A 2010 *J. Am. Chem. Soc.* **132** 13000

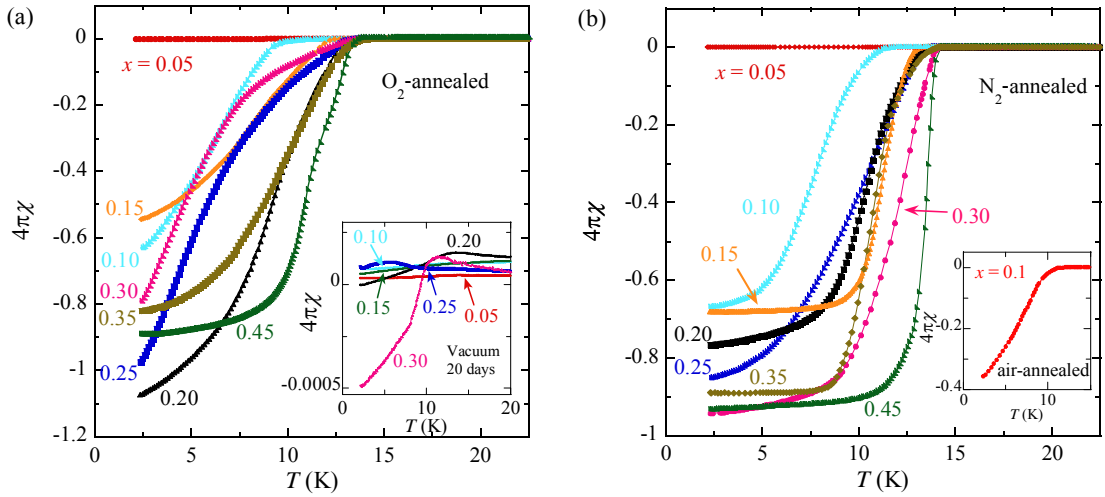

**Figure 1.** dc magnetic susceptibility as a function of temperature for the $Fe_{1+y}(Te_{1-x}Se_x)$ samples annealed in (a) oxygen and (b) nitrogen. Inset of (a): Susceptibility of the samples annealed in vacuum annealing for 20 days. Inset of (b): Susceptibility of the $x = 0.1$ sample annealed in air at 270 °C for 2 hrs. All samples were measured under magnetic field of 30 Oe with zero-field cooling histories.

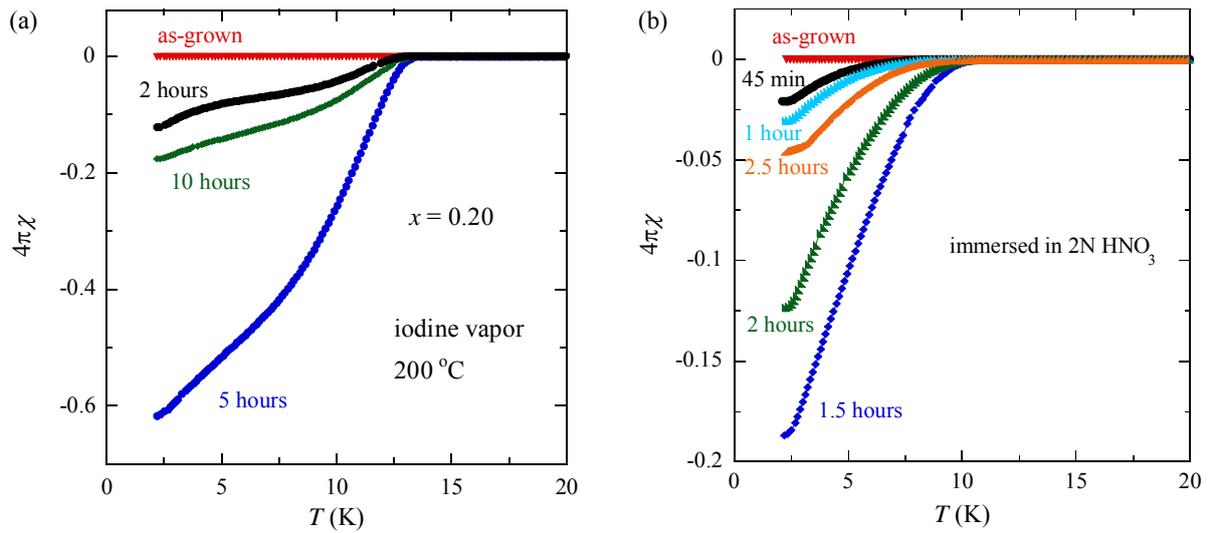

**Figure 2.** Magnetic susceptibility as a function of temperature for the $x = 0.2$ samples (a) annealed in iodine vapor and (b) treated with 2 N nitric acid for various time periods (see text).

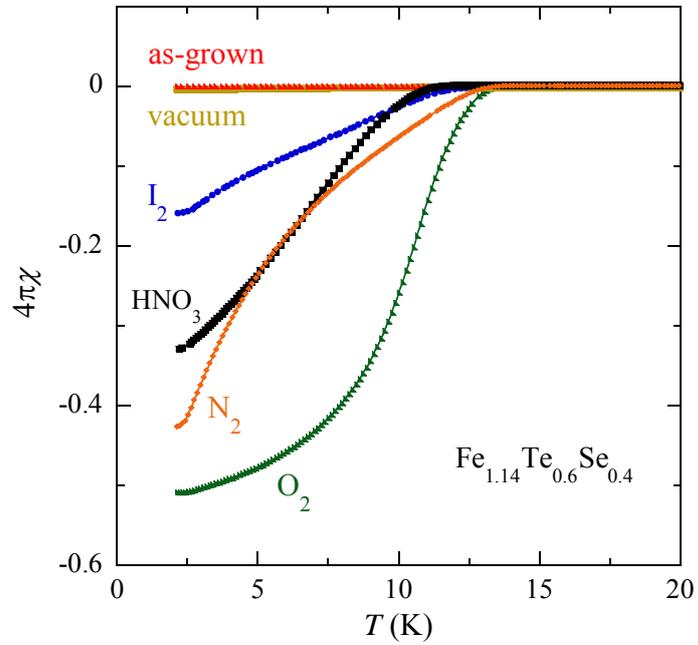

**Figure 3.** Magnetic susceptibility as a function of temperature for the as-grown, annealed (including vacuum, $O_2$-, $N_2$- and $I_2$ vapor annealing), and $HNO_3$-treated $Fe_{1.14}Te_{0.6}Se_{0.4}$ samples. $O_2$- and $N_2$-annealing were performed at 300 °C for 10 hrs, while $I_2$-annealing was conducted at 200 °C for 10 hrs to avoid sample degradation. The $HNO_3$ treatment was preceded by immersing the sample in 2N $HNO_3$ solution for 1.5 hrs.

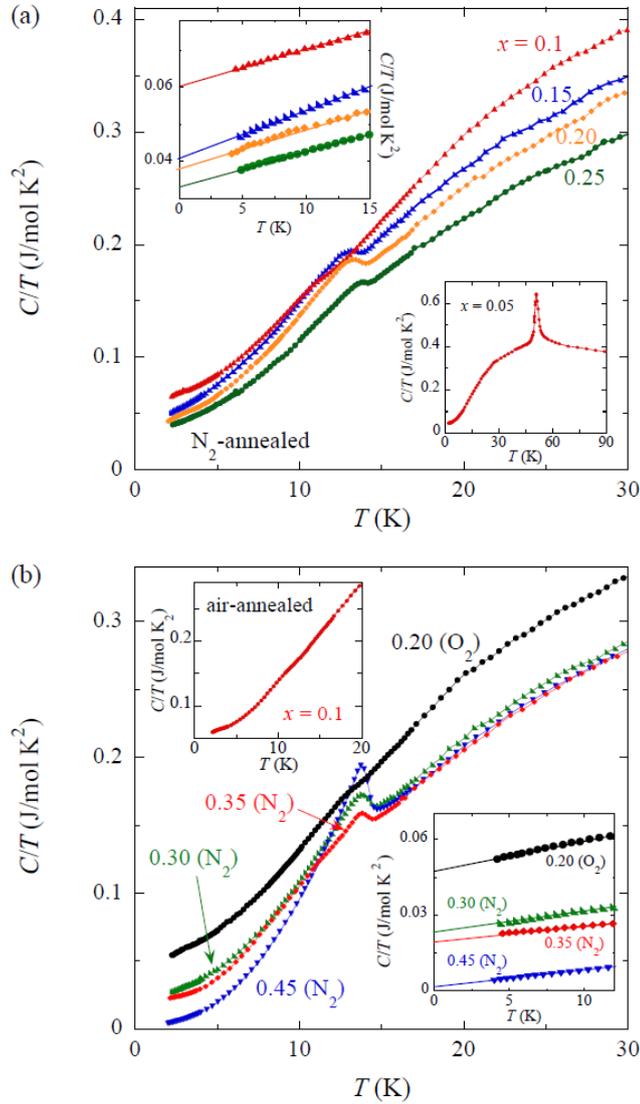

**Figure 4.** Specific heat divided by temperature $C/T$ as a function of temperature for $N_2$-annealed $Fe_{1+y}(Te_{1-x}Se_x)$ samples with $x = 0.1$-$0.25$ (a) and $x = 0.30 – 0.45$ (b). The specific heat data of the $O_2$-annealed, $x = 0.2$ sample is also shown in (b). Left inset in (a): linear fit for $C/T$ vs. $T^2$ at low temperatures; Right inset in (a): specific heat data for the $x = 0.05$ sample. Left inset in (b): specific heat data of the $x = 0.1$ sample annealed in air; Right inset in (b): linear fit for $C/T$ vs. $T^2$ at low temperatures.

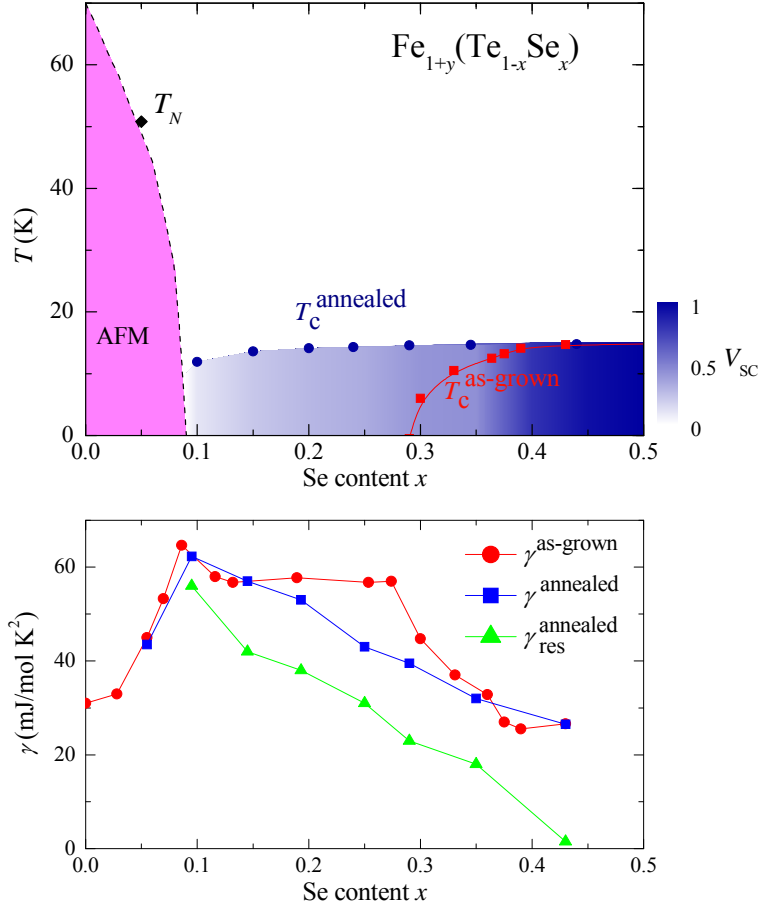

**Figure 5**. (a) The magnetic and electronic phase diagram for as-grown and $N_2$-annealed $Fe_{1+y}(Te_{1-x}Se_x)$ ($0 \leq x < 0.5$) sample. The data for Néel temperature $T_N$ (dashed line) and the bulk superconducting transition temperature $T_c^{as\text{-}grown}$ (red solid squares) for as-grown samples are quoted from our previous reports [1-2], while the superconducting transition temperatures of the annealed samples, $T_c^{annealed}$, are determined by the superconducting anomaly peaks in specific heat. The contour plot illustrates the evolution of the superconducting volume fraction $V_{SC}$ estimated from the specific heat data shown

in Figure 4. (b) Sommerfeld coefficient $\gamma^{\text{annealed}}$ and residual electronic specific heat coefficient $\gamma_{\text{res}}^{\text{annealed}}$ as a function of Se content $x$ for the $N_2$-annealed samples. The Sommerfeld coefficients of the as-grown samples $\gamma^{\text{as-grown}}$ (quoted from ref. [2]) are included here for comparison.